\documentclass[usenatbib,letters]{mnras}
\usepackage{newtxtext,newtxmath}
\usepackage[T1]{fontenc}

\DeclareRobustCommand{\VAN}[3]{#2}
\let\VANthebibliography\thebibliography
\def\thebibliography{\DeclareRobustCommand{\VAN}[3]{##3}\VANthebibliography}

\usepackage{amsmath}
\usepackage{graphicx}
\usepackage[usenames,dvipsnames]{xcolor}
\usepackage{mathtools}
\usepackage{gensymb}
\usepackage{tabularx}
\usepackage{subfigure}
\usepackage{physics}
\usepackage{subfiles}

\newcommand{\dobib}{\bibliography{references}}
\newcommand{\Da}{Damk{\"o}hler }

\newcommand{\civ}{\ion{C}{iv}}
\newcommand{\nv}{\ion{N}{v}}
\newcommand{\siiv}{\ion{Si}{iv}}
\newcommand{\ovi}{\ion{O}{vi}}
\newcommand{\oiii}{\ion{O}{iii}}

\title{A Model for Line Absorption and Emission from Turbulent Mixing Layers}

\author[B. Tan, S.P. Oh] {
  Brent Tan$^1$\thanks{E-mail: zunyibrent@physics.ucsb.edu},
  and S. Peng Oh$^1$\\
    $^1$University of California - Santa Barbara,
    Department of Physics, CA 93106-9530, USA\\
}
\date{Accepted XXX. Received YYY; in original form ZZZ}
\pubyear{2021}

\begin{document}
\renewcommand{\dobib}{}
\label{firstpage}
\pagerange{\pageref{firstpage}--\pageref{lastpage}}
\maketitle
\begin{abstract}
Turbulent mixing layers (TMLs) are ubiquitous in multiphase gas. They can potentially explain observations of high ions such as \ovi, which have significant observed column densities despite short cooling times. Previously, we showed that global mass, momentum and energy transfer between phases mediated by TMLs is not sensitive to details of thermal conduction or numerical resolution. By contrast, we show here that observables such as temperature distributions, column densities and line ratios {\it are} sensitive to such considerations. We explain the reason for this difference. We develop a prescription for applying a simple 1D conductive-cooling front model which quantitatively reproduces 3D hydrodynamic simulation results for column densities and line ratios, even when the TML has a complex fractal structure. This enables sub-grid absorption and emission line predictions in large scale simulations. The predicted line ratios are in good agreement with observations, while observed column densities require numerous mixing layers to be pierced along a line of sight. 
\end{abstract}

\begin{keywords}
% choose six from https://academic.oup.com/DocumentLibrary/mnras/keywords.pdf
hydrodynamics -- instabilities -- turbulence -- galaxies: haloes -- galaxies: clusters: general -- galaxies: evolution
\vspace{-1.5em}
\end{keywords}

%%%%%%%%%%%%%%%%%%%%%%%%%%%%%%%%%%%%%%%%%%%%%%%%%%%%%%%%%%%%%%%%%%%%%%%%%%%%%%%%%%
%                                   Body                                         %
%%%%%%%%%%%%%%%%%%%%%%%%%%%%%%%%%%%%%%%%%%%%%%%%%%%%%%%%%%%%%%%%%%%%%%%%%%%%%%%%%%

\section{Introduction} \label{sect:intro}

Observations of ultraviolet absorption lines of high ions like \civ, \siiv, \nv~and~\ovi~trace intermediate temperature ($\sim 10^5$~K) gas, assuming collisional ionization equilibrium (CIE). They are widely observed in a range of astrophysical contexts, such as our own galactic disk and halo, external galaxies, and high velocity clouds. Observations of a significant column density of these ions is puzzling since the gas they trace should cool quickly. One possibility is that they exist in turbulent mixing layers (TMLs) between cold $10^4$~K and hot $10^6$~K gas. There, radiative cooling is balanced by enthalpy flux into the TML, reaching a steady state. Such TMLs are relevant to a host of issues such as the stability and survival of AGN jets \citep{hardee97}, cold clouds in a hot wind \citep{scannapieco15,schneider17,gronke18,gronke20}, and cold streams inflowing from cosmological accretion \citep{mandelker20}. Observational diagnostics of TMLs could be very informative. For instance, the mass entrainment rate per unit area is directly proportional to the bolometric surface brightness (in the absence of radiative heating and scattering).

Models for absorption and emission in conductive-cooling fronts \citep{mckee77,borkowski90,gnat10} have predicted column densities that are too low, requiring many layers to be pierced along a sightline, and line ratios that do not match observations \citep{wakker12}. To our knowledge, the only equivalent work for TMLs is \citet{slavin93}, based on an analytic model by \citet{begelman90}. The latter had many important physical insights, but its detailed predictions are not in agreement with 3D hydrodynamic simulations \citep{ji19}. However, \citet{slavin93} made the important early prediction that TML column densities are also too low. This has been borne out in simulations \citep{kwak10,ji19}, although \citet{kwak10} find line ratios in good agreement with observations \citep{wakker12}. Subsequent modeling of TMLs \citep{ji19,fielding20,tan21} has focused on hot gas mass entrainment rates, which is crucial for the cold gas survival issues mentioned above. Such work has found that global mass, momentum and energy transfer between phases is not sensitive to details of thermal conduction or numerical resolution \citep{tan21}. By contrast, we show in this {\it Letter} that the same is not true of observables such as temperature distributions, column densities and line ratios. We develop a prescription for how we can apply 1D conductive-cooling front models to 3D TMLs to obtain analytic predictions for these quantities, which we then verify with simulations.

\dobib

\vspace{-30pt}

\section{Methods} \label{sect:methods}

We carry out our simulations using the publicly available MHD code Athena\verb!++! \citep{Stone2020}. For details of our setup and implementation, we refer the reader to Sections~2 and 5.1 of \citet{tan21}. In brief, we simulate a shear layer between gas of temperatures $T_{\rm cold} = 10^{4}$~K and $T_{\rm hot} = 10^{6}$~K, with shear velocity $v_{\rm shear} = 100 \, {\rm km \, s^{-1}}$, and include radiative cooling along with isotropic thermal conduction. We use a conductivity $\kappa_{\rm cond} = T_6^{\alpha} 10^6 {\rm ~erg~cm^{-1} ~s^{-1} ~K^{-1}}$, where $T_6 = T/10^6$~K, and a CIE cooling function $\Lambda$ based on a broken power law fit to \citet{Gnat_2007}. When we vary conduction and cooling in our simulations, we label them as $\kappa_n$ and $\Lambda_n$ respectively, where the subscript $n$ denotes a constant prefactor multiplying the fiducial values as stated above. We also label $\alpha_m$ where $m$ is the value of the exponent $\alpha$. (The fiducial simulation is thus $\kappa_1 \Lambda_1 \alpha_0$.) 

We calculate ion column densities along sightlines through simulations using Trident \citep{trident}, which generates synthetic spectra. For simplicity and consistency, we ignore photoionization and instead assume CIE ion fractions from \citet{Gnat_2007}. Using Trident, we add fields for the ions we are interested in by post-processing snapshots from simulation data. We assume solar metallicity and zero redshift. Lastly, we use pyatomdb \citep{pyatomdb} to compute line emissivities.

% With photoionization, then column densities of some low ions become sensitive to the amount of $T=10^{4}$K gas in the box. In practice, the amount of post mixing layer photo-ionized gas is limited by self-shielding, but for simplicity we do not model this. 
\dobib

\vspace{-1.5em}

\section{1D Mixing Layer Models} \label{sect:model}

Why should complex TMLs be amenable to 1D modeling? Here we justify this approach. TMLs can be characterized by their \Da number, ${\rm Da}=t_{\rm turb}/t_{\rm cool}$, the ratio of the eddy turnover time of the largest eddies to the cooling time. While $t_{\rm cool}$ is temperature dependent, it has proven useful to evaluate Da at the temperature of mixed gas, $T_{\rm mix} \sim (T_{\rm cold} T_{\rm hot})^{1/2}$, which we henceforth assume. TMLs can be either single phase (${\rm Da} < 1$), with temperature varying continuously with depth in the interface, or multiphase (${\rm Da} > 1$), with the slice-averaged cold gas fraction changing continuously \citep{tan21}. If multiphase, the interface has a large scale fractal structure down to the scale of the interface width \citep{fielding20}. In both regimes, numerically converged mass entrainment rates do not require the resolution of the Field length or even the interface, only the outer eddy scale of turbulence in cold gas \citep{tan21}. This is of order the box size here, and the cold cloud size in driven turbulence \citep{gronke21}.

That the temperature distribution of a single phase TML can be reproduced by a 1D model is reasonable. But doing so for a fractal, multiphase TML might appear implausible. It is useful to distinguish between macroscopic and microscopic heat diffusion. Macroscopic heat diffusion (such as turbulence) drives global energy transport, dictating the global structure of the TML and the {\it coarse-grained} temperature profile\footnote{Indeed, \citet{tan21} showed that the mean, coarse-grained temperature profile $\bar{T}(x) \approx f_c(x) T_{\rm cold} + (1-f_c(x)) T_{\rm hot}$, where $f_c(x)$ is the spatially varying mass fraction of cold gas, can be reproduced by a mixing length model.}. However, it does not drive actual energy exchange between the two phases; all that changes is the relative amount of hot and cold gas. Changing the {\it fine-grained}, thermodynamic temperature -- which determines ionic abundances -- requires microscopic heat transfer via explicit thermal conduction. Each segment of the fractal interface between hot and cold gas is a locally planar, laminar heat conduction front, whose temperature profile is set by the competition between explicit thermal conduction and cooling. All intermediate temperature gas lies in this interface, which {\it can} be modeled in 1D. Since the same universal interface profile holds at every segment of the fractal interface, it sets the temperature PDF. The global structure of the TML is immaterial. In \S\ref{sect:sims}, we verify this conjecture by comparing 1D models to 3D simulations. 

Thus, if one is to accurately capture this temperature distribution in 3D simulations, then in contrast to mass entrainment, the interface (and hence the Field length) {\it must} be resolved. This is currently impossible in large-scale simulations. Moreover, explicit thermal conduction {\it must} be included. 

\subsection{1D Mixing Layers}

In a 1D mixing layer, an equilibrium state can be reached between two stable phases with three ingredients: radiative cooling, thermal conduction, and enthalpy advection. In the frame of the front, this gives \citep{kim}:
\begin{align}
    \frac{{\rm d}}{\dd x} \left( \kappa \frac{\dd T}{\dd x} \right) &= j_x c_p \frac{\dd T}{\dd x} + \rho \mathcal{L}(T) ,
    \label{eqn:ODE} 
\end{align}
where $j_x = \rho v_x$ is the the constant mass flux and $c_p = \frac{\gamma}{\gamma-1} \frac{k_{\rm B}}{\bar{m}}$ is the specific heat at constant pressure. $\kappa$ is the thermal conductivity and $\rho \mathcal{L} = n^2 \Lambda - n\Gamma$ is the net cooling rate per unit volume, where $\Lambda$ is the cooling function and $\Gamma$ is the heating rate. We assume that $\rho v^2 \ll P$ so pressure is constant, which can be verified in the solutions. Given the boundary conditions $T_{-\infty} = T_{\rm cold}$, $T_{\infty} = T_{\rm hot}$ and $\frac{\dd T}{\dd x}_{\pm \infty} = 0$, we can solve for the equilibrium solution numerically with $j_x$ as an eigenvalue using the shooting method. We can also integrate Eq.~\eqref{eqn:ODE} to give us the relationship between $j_x$ and the surface brightness $Q$:
\begin{align}
    j_x = \frac{Q}{c_p (T_{\rm hot}-T_{\rm cold})}; \ \ Q = -\int^{\infty}_{-\infty} \rho \mathcal{L} \, \dd x.
    \label{eqn:mflx}
\end{align}

\subsection{Temperature Distribution}

\begin{figure} 
    \centering
    \includegraphics[width=\columnwidth]{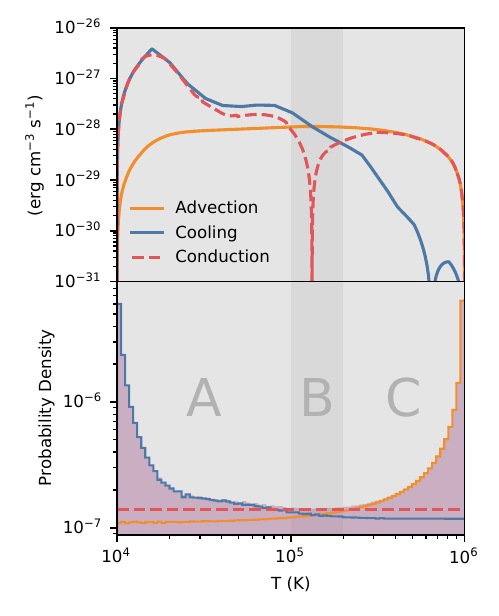}
    \vspace{-3.0em}
    \caption{{\it Lower panel:} The lilac histogram shows the temperature distribution across the front. Three separate regions are identified where the distribution can be understood by simplifying Eq.~\eqref{eqn:ODE}, with colored curves showing resulting distributions. {\it Upper panel:} Corresponding magnitudes of the advection, cooling and conduction terms in Eq.~\eqref{eqn:ODE}.}
    \label{fig:distribution}
    \vspace{-1.0em}
\end{figure}

For a given front solution $T(x)$, the volume weighted probability density function (PDF) of the temperature distribution is given by $\dd x/\dd T$ multiplied by some normalization factor, henceforth referred to in our plots as just the probability density. In the lower panel of Fig.~\ref{fig:distribution}, we show in the shaded lilac histogram the temperature distribution for the solution to Eq.~\eqref{eqn:ODE} with our fiducial parameters. We consider temperatures in the range from $10^4$ to $10^6$~K, excluding the boundary temperatures themselves. In the upper panel, we show the corresponding magnitudes of the advection, cooling and conduction terms in Eq.~\eqref{eqn:ODE}. Similar to the analysis in \citet{mckee77} for spherical clouds, we identify three separate regions where the distribution can be understood via simplified versions of Eq.~\eqref{eqn:ODE}.

\begin{itemize}
    \item {\it Region A} : In the low temperature region, we can see from the upper panel of Fig.~\ref{fig:distribution} that cooling dominates over advection and is balanced by the conduction term. Eq.~\eqref{eqn:ODE} thus simplifies to
     \begin{equation}
        \frac{{\rm d}}{\dd x} \left( \kappa \frac{\dd T}{\dd x} \right) = \rho \mathcal{L}.
     \label{eqn:region_A} 
     \end{equation}
     The blue curve in the lower panel shows the temperature distribution of the solution to this simplified equation, in excellent agreement with the actual distribution in this region (normalizations of the colored lines in the lower panel have been adjusted for easy comparison). From Eq.~\eqref{eqn:region_A}, we can define a characteristic length scale known as the Field length \citep{mnb},
     \begin{align}
        \lambda_{\rm F} = \sqrt{\frac{\kappa T}{n^2 \Lambda}}.
     \label{eqn:field_length} 
     \end{align}
                     
    \item {\it Region B} : At some intermediate temperature, the cooling and advection terms are equal and the conduction term is zero, hence: 
     \begin{align}
         \kappa \frac{\dd T}{\dd x} \approx \text{constant}.
     \label{eqn:region_B} 
     \end{align}
     This is a single inflection point separating regions A and C. 
     % The distribution here hence has the same scaling with temperature as $\kappa$. For example, the red dashed line in the lower panel of Fig.~\ref{fig:distribution} is horizontal since $\kappa$ is constant. The temperature where this happens increases when kappa has a stronger scaling with T.

    \item {\it Region C} : In the high temperature region where cooling is weak, the advection term dominates and is balanced by the the conduction term. Eq.~\eqref{eqn:ODE} thus simplifies to 
     \begin{equation}
        \frac{{\rm d}}{\dd x} \left( \kappa \frac{\dd T}{\dd x} \right) = j_x c_p \frac{\dd T}{\dd x}.
     \label{eqn:region_C} 
     \end{equation}
     As with Region A, the orange curve in the lower panel of Fig.~\ref{fig:distribution} shows the temperature distribution of the solution to Eq.~\eqref{eqn:region_C}. In fact, for a constant $\kappa$, we can solve Eq.~\eqref{eqn:region_C} analytically, which gives an exponential temperature profile with a distribution that scales as $\lambda_{\rm D}/(T_{\rm hot} - T)$, where $\lambda_{\rm D}$ is a diffusion length scale: 
     \begin{equation}
        \lambda_{\rm D} = \frac{\kappa}{j_x c_p}.
     \label{eqn:diffusion_length} 
     \end{equation}
    % It is worth noting that the characteristic length scales $\lambda_{\rm F},\lambda_{\rm D}$  are not the same as the scaling of the probability density with temperature, which are instead dx/dt

\end{itemize}
% We can also see from this that the heating function doesn't affect the energy flux ... but prevents the hot phase from cooling.

\subsection{Length Scales} \label{subsect:length_scaling}
\begin{figure} 
    \centering
    \includegraphics[width=\columnwidth]{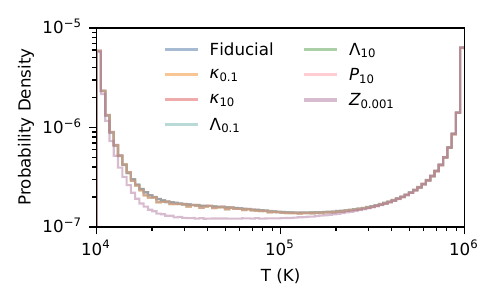}
    \vspace{-3.0em}
    \caption{Temperature distributions remain unchanged for different values of $\kappa$, $\Lambda$ and $P$ relative to fiducial values. Subscripts denote multiples of the fiducial value. However, changing the metallicty can change the distribution by changing the shape of the cooling curve.}
    \label{fig:magnitude}
    \vspace{-1.0em}
\end{figure}
In regions A and C, the temperature scale heights $\lambda_{\rm F},\lambda_{\rm D}$ are obtained by balancing conduction against cooling and enthalpy advection respectively (i.e., the second and third term of Eq.~\eqref{eqn:ODE}). Since there is no net heat flux into the TML ($\frac{\dd T}{\dd x}_{\pm \infty} = 0$), the latter two balance across the front as a whole, i.e. enthalpy advection balances cooling, giving Eq.~\eqref{eqn:mflx}, which gives $j_x \propto Q \propto (n^{2} \Lambda) \lambda$ at a given temperature. Substituting into Eq.~\eqref{eqn:diffusion_length}, this gives $\lambda_{\rm D} \propto \sqrt{\kappa_0/(n^{2} \Lambda_0)} \propto \lambda_{\rm F}$, where $\kappa_0$ and $\Lambda_0$ are constant prefactors multiplying $\kappa (T)$ and $\Lambda(T)$, i.e. both the Field length $\lambda_{\rm F}$ (Eq.~\eqref{eqn:field_length}) and the diffusion length $\lambda_{\rm D}$ (Eq.~\eqref{eqn:diffusion_length}) share the same scalings with respect to $\kappa_0$, $\Lambda_0$, and $P \propto n$ (at fixed $T$). Since varying any of these rescales the solutions in regions A and C identically, the temperature distribution ($\propto dx/dT$) is thus {\it independent} of $\kappa_0$, $\Lambda_0$ and $P$, unless they change Da sufficiently to affect the choice of $\kappa$ (see \S\ref{subsect:alphas}). This is verified numerically in Fig.~\ref{fig:magnitude}, where we show that the distribution remains unchanged whether we vary $\kappa_0$, $\Lambda_0$ or $P$. Thus, a change in isobaric cooling time or reduced conduction due to tangled B-fields does not affect the temperature PDF. While changes in the normalization of cooling or conduction processes do not affect the temperature PDF, changes in their temperature dependence (e.g., via metallicity for cooling) do, as we now discuss. 

\subsection{Non-Constant Conductivity} \label{subsect:alphas}
Consider a temperature dependent conductivity $\kappa \propto T^{\alpha}$. What are relevant values of $\alpha$? For single phase TMLs (${\rm Da} < 1$), where the coarse-grained and fine-grained temperatures coincide, the temperature PDF is set by turbulent heat diffusion:
\begin{itemize} 
 \item $\alpha=-0.5$ : Since conductivity and diffusivity $D \sim v L$ are related by $\kappa = D \rho c_p \propto D P/T$, this arises when $D \propto T^{0.5}$. This is seen in the low Da regime of TMLs, where turbulent diffusion scales with the local sound speed, $D_{\rm turb} \propto c_{s} \propto T^{0.5}$ (see Figure 14 of \citealp{tan21}).
 \end{itemize} 
For multiphase TMLs (${\rm Da} > 1$), the temperature PDF is set by microscopic thermal conduction:
\begin{itemize}
    \item $\alpha=2.5$ : Spitzer conductivity. \cite{spitzer} gives the thermal conductivity of an ionized plasma as: 
    \begin{align}
        \kappa_{\rm sp} &= 5.7 \times 10^{-7} \; T^{2.5} \; \text{erg cm$^{-1}$ s$^{-1}$ K$^{-1}$} .
    \end{align}
    
    \item $\alpha=0$ : Constant conductivity. This was assumed in previous simulations of TMLs which included thermal conduction (e.g. \citealp{kim,tan21}), largely for numerical reasons. 

    \item $\alpha=-1$ : Constant diffusivity. This is a good approximation for numerical diffusion $D_{\rm num} \sim v \Delta x$ in simulations without thermal conduction (e.g. \citealp{kwak10,ji19}).
    
\end{itemize}

\begin{figure} 
    \centering
    \includegraphics[width=\columnwidth]{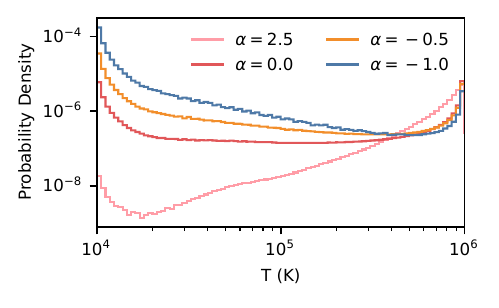}
    \vspace{-3.0em}
    \caption{Distributions for various values of $\alpha$. The scaling with temperature gets steeper as $\alpha$ increases.}
    \label{fig:pdfs}
    \vspace{-1.0em}
\end{figure}

In Fig.~\ref{fig:pdfs}, we show the corresponding temperature distributions for each $\alpha$. As $\alpha$ increases, the temperature scale height and PDF is weighted toward higher temperatures. 

\subsection{Column Densities and Line Ratios}
Given a solution for the temperature profile, we can compute column densities. From the scalings in \S\ref{subsect:length_scaling}, and using $N_{\ovi} = \int n_{\ovi} \, \dd l \propto n_{\ovi} \lambda$, we obtain: 

\begin{align}
    N_{\ovi} = \ell N_c
    \left(\frac{\kappa}{10^6 \,{\rm erg \, cm^{-1} s^{-1} K^{-1}}} \right)^{0.5}
    \left(\frac{\Lambda(T_{\ovi})}{10^{-21.5}\, {\rm erg \, cm^{3} s^{-1}}}  \right)^{-0.5} 
    \left(\frac{Z}{Z_{\odot}}\right),
    \label{eqn:col_dens_ovi}
\end{align}
where $N_c$ is a reference column density, $\kappa$ is the conductivity, $\Lambda(T_{\ovi})$ is the the cooling function at $T_{\ovi} \sim 3 \times 10^5$~K (where \ovi~abundance peaks), and $\ell$ is a correction factor. Similar equations can be written for the other ions as well. Since $\lambda_{F}, \lambda_{D} \propto \sqrt{\kappa/(n^{2} \Lambda(T))}$ have the same scalings, the scaling $N_{i} \propto \sqrt{\kappa/\Lambda}$ holds whether the ion peaks in region A or C. The value $N_c$ depends on $\alpha$. For example, in a turbulent single phase front where $\alpha=-0.5$, $N_c = 5.10 \times 10^{11}$~cm$^{-2}$, but in a front with just Spitzer conduction where $\alpha=2.5$, $N_c = 2.1 \times 10^{12}$~cm$^{-2}$. $N_c$ also depends on the shape of the cooling curve (and hence metallicity indirectly). 

We include a path length correction factor $\ell$ to match 3D simulations, since sight-lines that intersect the mixing layer at an angle have longer path lengths. We estimate this to be a factor of $\sim \sqrt{2}$. This correction factor could also account for a sightline intersecting the interface multiple times in a fractal TML. However, because the mixing layer does not often `wrap around' on large scales, we find in our 3D simulations that the sightlines usually only intersect the mixing layer 1-2 times. Similarly, a line through the fractal coastline on a map will typically intersect the water-land boundary once. 

An important point in applying Eq.~\eqref{eqn:col_dens_ovi} is the choice of $\kappa$. The two candidates are explicit thermal conduction $\kappa_{\rm cond}$ and turbulent conduction $\kappa_{\rm turb}$. If $\kappa_{\rm cond} > \kappa_{\rm turb}$, then $\kappa_{\rm cond}$ should be used. However if $\kappa_{\rm cond} < \kappa_{\rm turb}$, then we have to consider Da of the system. For large Da (multiphase), $\kappa_{\rm cond}$ should be used, since the width of individual interfaces are governed by explicit thermal conduction. If ${\rm Da} < 1$ (single phase), they are set by turbulent conduction, and $\kappa_{\rm turb}$ should be used instead.

The scalings of Eq.~\eqref{eqn:col_dens_ovi} are consistent with those of Equation~30 in \citet{ji19}. They found that $N_{\ovi} \propto Z^{0.8}$, but with the change in cooling function with metallicity folded in. In the single phase regime that their simulations fall in, $\kappa_{\rm turb} = D_{\rm turb} \rho c_p \propto P$ which then translates to $N_{\ovi} \propto P^{0.5}$, as seen in their simulations.

Similarly, we can also compute a line surface brightness $Q_i$ as
\begin{align}
    Q_{i} = f_{i} Q_{\rm total} \ \ ; \ \  f_{i} \equiv \frac{\int n^2 \epsilon_{i}(T) \, \dd l}{\int n^2 \Lambda (T) \, \dd l},
\label{eqn:qion}
\end{align}
where $\epsilon_{i}$ is the line emissivity and $Q_{\rm total}$ is the total surface brightness modeled in \citet{tan21} (which for a fractal interface differs from Eq.~\ref{eqn:mflx}). 
For example, using pyatomdb to compute the emissivity of [\oiii] 5008.24 \AA, we obtain $f_i \sim 8.4 \times 10^{-3}$ for $\alpha = 0$ and $f_i \sim 5.2 \times 10^{-3}$ for $\alpha = -0.5$.

\dobib

\vspace{-1.5em}

\section{Results} \label{sect:sims}
\subsection{Temperature PDFs}
\begin{figure} 
    \centering
    \includegraphics[width=\columnwidth]{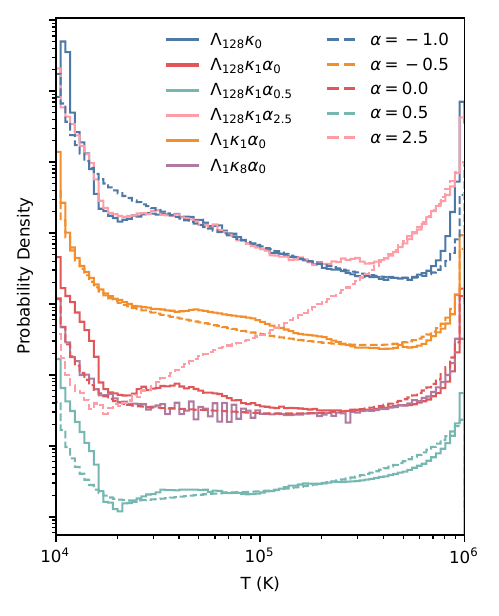}
    \vspace{-3.0em}
    \caption{Distributions of simulations (solid lines) compared to expected corresponding 1D models (dashed lines). Simulations vary cooling and conduction prefactors, along with $\alpha$. Normalizations are adjusted for comparison.}
    \label{fig:distributions_sim}
    \vspace{-1.0em}
\end{figure}
In Fig.~\ref{fig:distributions_sim}, we compare 3D hydrodynamic simulations with 1D models. The distributions are from single snapshots where the mixing layer has fully developed. The following categories are represented:
\begin{itemize}
\item {\it Strong Conduction}: When $\kappa_{\rm cond} > \kappa_{\rm turb}$ (purple line $\Lambda_1 \kappa_8 \alpha_0$), mixing is set by conduction and not by turbulence, and hence the distribution follows the constant $\kappa$ model ($\alpha=0$, red dashed line). 

\item {\it Single Phase}: In the low Da regime, where mixing is faster than cooling, the gas is single phased. Our fiducial setup ($\Lambda_1 \kappa_1 \alpha_0$, orange line) lies in this region. Since $\kappa_{\rm turb}$ is larger than $\kappa_{\rm cond}$ and hence dominates mixing, we expect the distribution to follow the $\alpha=-0.5$ model (orange dashed line), as explained in \S\ref{subsect:alphas}.

\item {\it Multiphase}: The rest of the simulations have strong cooling ($\Lambda_{128}$) and hence lie in the high Da regime. Although $\kappa_{\rm turb}$ is larger than $\kappa_{\rm cond}$, the multiphase structure of the mixing layer means that the thickness of the interface locally is still set by $\kappa_{\rm cond}$. We show simulations for a range of $\alpha$, including one with no explicit conduction (blue line) which hence only has numerical diffusivity ($\alpha=-1$, blue dashed line). The simulation with a Spitzer scaling (pink line) differs from the $\alpha=2.5$ model for $T < 3 \times 10^{5}$~K. This is because the Field length at lower temperatures is unresolved, as we now discuss. 
\end{itemize}

\vspace{-1.5em}
\subsection{Resolution}
What resolution is required for convergence? It is usually thought that one needs to resolve the Field length $\lambda_{\rm F}$ (e.g. see Figure 7 of \citealp{kim}). Our results are consistent with this. In our highest resolution simulations, we are just able to resolve the smallest $\lambda_{\rm F}$ with fiducial cooling ($\Lambda_1$). However, in simulations with strong cooling ($\Lambda_{128}$), the Field length $\lambda_{\rm F}$ of gas below $T=5 \times 10^4$~K remains unresolved. As a result, we see in Fig.~\ref{fig:distributions_sim} that these simulations all show a dip at $\sim 2 \times 10^4$~K, where $\lambda_{\rm F}$ is the smallest.
In resolution tests, this feature becomes more prominent as we lower the resolution. However, if the lines we are interested in only trace gas at $T \sim 10^5$~K, then it is sufficient to just have enough resolution to resolve $\lambda_{\rm F}$ at $10^5$~K. The lowest resolution simulation also shows a drop at higher temperatures, as numerical diffusion starts to dominate over thermal conduction. This can also be seen in the simulation with Spitzer conductivity (pink line in  Fig.~\ref{fig:distributions_sim}), where the distribution switches over from the expected model to the one with numerical diffusion at lower temperatures where the Spitzer conductivity is small.

\vspace{-1.5em}
\subsection{Column Densities and Line Ratios}
\begin{figure} 
    \centering
    \includegraphics[width=\columnwidth]{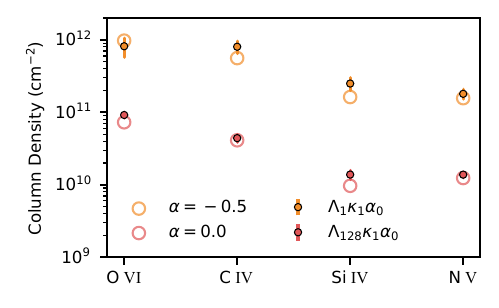}
    \vspace{-3.0em}
    \caption{Column densities from simulation sightlines (solid circles) compared with their corresponding $\alpha$ models (hollow circles).}
    \label{fig:col_dens}
    \vspace{-1.0em}
\end{figure}
Finally, we use Trident to compute column densities along sightlines through the mixing layer simulations. We run a hundred sightlines through the mixing layer in the simulation, randomly initializing the start and end points of the sightlines on the sides of the box each time. These column densities are then summed. While there can be a large variation in column densities along a single sightline, this is greatly reduced when passing through many mixing layers, as required to match observed column densities. To estimate the variance, we repeat this process 75 times over several time snapshots.

The average column densities per mixing layer are plotted in Fig.~\ref{fig:col_dens} for the fiducial setup ($\Lambda_{1}$, single phase) and one with much stronger cooling ($\Lambda_{128}$, multiphase). Each setup is also compared to the model predictions from \S\ref{sect:model} with $\ell=\sqrt{2}$. For the $\alpha=-0.5$ model, we used $\kappa_{\rm turb} = 10^7 {\mathrm{~erg}\,\mathrm{~cm}^{-1}\,\mathrm{s}^{-1}\,\mathrm{K}^{-1}} (T/10^4\mathrm{~K})^{-0.5}$ from Figure~14 of \citet{tan21}. The models and simulations are in good agreement. This implies that the number of interface intersections per mixing layer in the multiphase regime is of order unity.

\begin{figure} 
    \centering
    \includegraphics[width=\columnwidth]{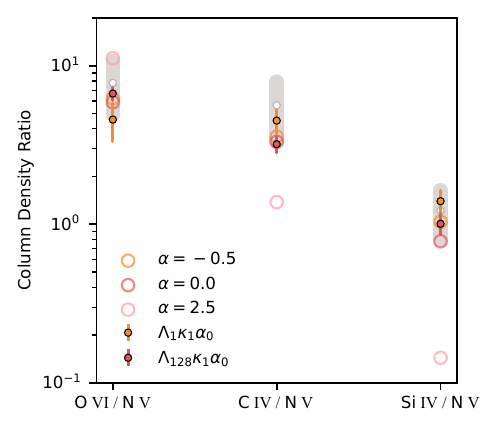}
    \vspace{-3.0em}
    \caption{Line ratios from simulations (solid circles) and their corresponding $\alpha$ models (hollow circles). Grey bars represent observational data of the MW halo from \citet{wakker12}.}
    \label{fig:lines_sim}
    \vspace{-1.0em}
\end{figure}

We show the corresponding model and simulation line ratios in Fig.~\ref{fig:lines_sim}. Observational values for the Milky Way were obtained from \citet{wakker12} and are also shown for comparison. We find that the single phase TML model is a good match with the observations, consistent with the 2D simulations of \citet{kwak10}. While the multiphase TML simulation is also a good match, the constant conductivity used therein is not physically motivated, unless conduction is saturated \citep{cowie77}. Assuming a Spitzer conductivity (pink hollow circles) instead leads to predictions which differ significantly. Such potential constraints on ${\rm Da} = t_{\rm turb}/t_{\rm cool} < 1$ are interesting, particularly if one of these timescales can be independently estimated. 

We also estimate line emission from the simulations above using pyatomdb and compare with Eq.~\ref{eqn:qion}. For [\oiii] 5008.24 \AA, we find $Q_i \sim 4 \times 10^{-10}$ erg cm$^{-2}$ s$^{-1}$ and $Q_i \sim 4.5 \times 10^{-9}$ erg cm$^{-2}$ s$^{-1}$ for $\Lambda_1$ and $\Lambda_{128}$ respectively. While these values are around a factor of 2 higher than the model values, the difference mostly comes from post-processing the line emission rather than tracking total emission in the simulations over a timestep as done for $Q_{\rm total}$.

\dobib

\vspace{-1.5em}

\section{Discussion} \label{sect:disc}

We have found that the thermodynamic temperature distribution in TMLs depends only on the local rather than global front structure. As shown in \citet{fielding20} and \citet{tan21}, the global structure is set by turbulence, which can significantly deform the front in the fast cooling limit, leading to a complicated fractal structure. The interaction between turbulence and cooling sets the overall surface brightness and mass entrainment rate. It is not sensitive to the details of thermal diffusion, and only requires the outer eddy scale to be resolved. However, the local front structure is set by balance between radiative cooling, advection, and thermal diffusion (conduction, turbulence), requiring much smaller scales such as the Field length to be resolved. Fortunately, the temperature PDFs, column densities and line ratios in converged 3D simulations are surprisingly well matched by simple 1D models of local front structure. This is good news, because resolving the local front structure is currently impossible in larger galaxy scale simulations. Instead, the model in this {\it Letter} can be inserted as a subgrid prescription, when calculating the contribution of TMLs to line emission or absorption. 

We regard this as significant progress. At the same time, there are refinements we have ignored, which we leave to future work. Some of these include: 
\begin{itemize} 
\item{{\it Photoionization; Non-equilibrium Ionization (NEI).} The effects of photoionization and NEI were studied in 3D simulations by \citet{ji19}. They have two effects: (i) the gas is over-ionized compared to CIE expectations, and thus has larger line column densities (typically by a factor of a few). (ii) Decreased radiative cooling efficiency due to  over-ionization; leading to thicker mixing layers and larger column densities, although in practice this is a minor effect. Both of these effects can be captured in our analytic model, by altering the temperature dependence of the ionization fraction $x_{i}(T)$ and cooling function $\Lambda(T)$, as a function of radiation field \citep{wiersma09} or cooling history \cite{Gnat_2007}, using lookup tables.} 
\item{{\it Multiple clouds; Kinematic Structure.} To account for the observations, a line of sight has to pass through $\sim 100-1000$ mixing layers; which is conceivable if the cold gas has a `fog-like' structure \citep{mccourt18,gronke20_fog}. Cold gas in a turbulent medium acquires a wide-ranging, almost scale-free range of sizes, but the covering fraction is dominated by small cloudlets \citep{gronke21}. Our analytic model enables us to assign column densities and line ratios for these cloudlets, while kinematic structure due to turbulence can be obtained from the simulation.} 
\item{{\it Nonthermal Forces.} Our simulations are purely hydrodynamic. B-fields can suppress the Kelvin-Helmholtz instability, significantly reduce mass entrainment rates and column densities \cite{ji19}. Non-thermal pressure support from cosmic-rays has similar effects (Tan et al 2021, in preparation).}   
\item{{\it Metallicity and Dust Depletion.} We have assumed solar and equal cold/hot gas metallicities/abundances. The first assumption is easily modified. The second can be handled by modeling the relative cold/hot gas fraction in the mixing layer.}
\item{{\it Anisotropic Conduction.} In our simulations, we only model isotropic conduction. Anisotropic conduction along tangled B-fields potentially implies a reduction in $\kappa$. As long as $\kappa(T)$ can be calibrated from high resolution simulations, it can be used in our 1D model.} 
\end{itemize} 

\dobib

\section*{Acknowledgements}

We thank M. Gronke, L. Lancaster and the anonymous referee for helpful comments. We acknowledge support from NASA grants NNX17AK58G, 19-ATP19-0205, HST-AR- 15797.001-A, NSF grant AST-1911198 and XSEDE grant TG-AST180036. This research was supported in part by the National Science Foundation under Grant No. NSF PHY-1748958 to KITP, and made use of \texttt{yt} \citep{yt}.

\section*{Data Availability}

Data will be shared upon reasonable request to the authors.

\dobib

\newpage
\bibliography{references}

\begin{thebibliography}{}
\makeatletter
\relax
\def\mn@urlcharsother{\let\do\@makeother \do\$\do\&\do\#\do\^\do\_\do\%\do\~}
\def\mn@doi{\begingroup\mn@urlcharsother \@ifnextchar [ {\mn@doi@}
  {\mn@doi@[]}}
\def\mn@doi@[#1]#2{\def\@tempa{#1}\ifx\@tempa\@empty \href
  {http://dx.doi.org/#2} {doi:#2}\else \href {http://dx.doi.org/#2} {#1}\fi
  \endgroup}
\def\mn@eprint#1#2{\mn@eprint@#1:#2::\@nil}
\def\mn@eprint@arXiv#1{\href {http://arxiv.org/abs/#1} {{\tt arXiv:#1}}}
\def\mn@eprint@dblp#1{\href {http://dblp.uni-trier.de/rec/bibtex/#1.xml}
  {dblp:#1}}
\def\mn@eprint@#1:#2:#3:#4\@nil{\def\@tempa {#1}\def\@tempb {#2}\def\@tempc
  {#3}\ifx \@tempc \@empty \let \@tempc \@tempb \let \@tempb \@tempa \fi \ifx
  \@tempb \@empty \def\@tempb {arXiv}\fi \@ifundefined
  {mn@eprint@\@tempb}{\@tempb:\@tempc}{\expandafter \expandafter \csname
  mn@eprint@\@tempb\endcsname \expandafter{\@tempc}}}

\bibitem[\protect\citeauthoryear{{Begelman} \& {Fabian}}{{Begelman} \&
  {Fabian}}{1990}]{begelman90}
{Begelman} M.~C.,  {Fabian} A.~C.,  1990, \mnras, 244, 26P

\bibitem[\protect\citeauthoryear{{Begelman} \& {McKee}}{{Begelman} \&
  {McKee}}{1990}]{mnb}
{Begelman} M.~C.,  {McKee} C.~F.,  1990, \mn@doi [\apj] {10.1086/168994}, 358,
  375

\bibitem[\protect\citeauthoryear{{Borkowski}, {Balbus}  \&
  {Fristrom}}{{Borkowski} et~al.}{1990}]{borkowski90}
{Borkowski} K.~J.,  {Balbus} S.~A.,   {Fristrom} C.~C.,  1990, \mn@doi [\apj]
  {10.1086/168784}, \href
  {https://ui.adsabs.harvard.edu/abs/1990ApJ...355..501B} {355, 501}

\bibitem[\protect\citeauthoryear{{Cowie} \& {McKee}}{{Cowie} \&
  {McKee}}{1977}]{cowie77}
{Cowie} L.~L.,  {McKee} C.~F.,  1977, \mn@doi [\apj] {10.1086/154911}, \href
  {https://ui.adsabs.harvard.edu/abs/1977ApJ...211..135C} {211, 135}

\bibitem[\protect\citeauthoryear{{Fielding}, {Ostriker}, {Bryan}  \&
  {Jermyn}}{{Fielding} et~al.}{2020}]{fielding20}
{Fielding} D.~B.,  {Ostriker} E.~C.,  {Bryan} G.~L.,   {Jermyn} A.~S.,  2020,
  \mn@doi [\apjl] {10.3847/2041-8213/ab8d2c}, 894, L24

\bibitem[\protect\citeauthoryear{{Foster} \& {Heuer}}{{Foster} \&
  {Heuer}}{2020}]{pyatomdb}
{Foster} A.~R.,  {Heuer} K.,  2020, \mn@doi [Atoms] {10.3390/atoms8030049},
  \href {https://ui.adsabs.harvard.edu/abs/2020Atoms...8...49F} {8, 49}

\bibitem[\protect\citeauthoryear{Gnat \& Sternberg}{Gnat \&
  Sternberg}{2007}]{Gnat_2007}
Gnat O.,  Sternberg A.,  2007, \mn@doi [\apjs] {10.1086/509786}, 168, 213

\bibitem[\protect\citeauthoryear{{Gnat}, {Sternberg}  \& {McKee}}{{Gnat}
  et~al.}{2010}]{gnat10}
{Gnat} O.,  {Sternberg} A.,   {McKee} C.~F.,  2010, \mn@doi [\apj]
  {10.1088/0004-637X/718/2/1315}, \href
  {https://ui.adsabs.harvard.edu/abs/2010ApJ...718.1315G} {718, 1315}

\bibitem[\protect\citeauthoryear{{Gronke} \& {Oh}}{{Gronke} \&
  {Oh}}{2018}]{gronke18}
{Gronke} M.,  {Oh} S.~P.,  2018, \mn@doi [\mnras] {10.1093/mnrasl/sly131}

\bibitem[\protect\citeauthoryear{{Gronke} \& {Oh}}{{Gronke} \&
  {Oh}}{2020a}]{gronke20}
{Gronke} M.,  {Oh} S.~P.,  2020a, \mn@doi [\mnras] {10.1093/mnras/stz3332},
  492, 1970

\bibitem[\protect\citeauthoryear{{Gronke} \& {Oh}}{{Gronke} \&
  {Oh}}{2020b}]{gronke20_fog}
{Gronke} M.,  {Oh} S.~P.,  2020b, \mn@doi [\mnras] {10.1093/mnrasl/slaa033},
  \href {https://ui.adsabs.harvard.edu/abs/2020MNRAS.494L..27G} {494, L27}

\bibitem[\protect\citeauthoryear{{Gronke}, {Oh}, {Ji}  \& {Norman}}{{Gronke}
  et~al.}{2021}]{gronke21}
{Gronke} M.,  {Oh} S.~P.,  {Ji} S.,   {Norman} C.,  2021, arXiv e-prints, \href
  {https://ui.adsabs.harvard.edu/abs/2021arXiv210713012G} {p. arXiv:2107.13012}

\bibitem[\protect\citeauthoryear{{Hardee} \& {Stone}}{{Hardee} \&
  {Stone}}{1997}]{hardee97}
{Hardee} P.~E.,  {Stone} J.~M.,  1997, \mn@doi [\apj] {10.1086/304208}, 483,
  121

\bibitem[\protect\citeauthoryear{{Hummels}, {Smith}  \& {Silvia}}{{Hummels}
  et~al.}{2017}]{trident}
{Hummels} C.~B.,  {Smith} B.~D.,   {Silvia} D.~W.,  2017, \mn@doi [\apj]
  {10.3847/1538-4357/aa7e2d}, \href
  {http://adsabs.harvard.edu/abs/2017ApJ...847...59H} {847, 59}

\bibitem[\protect\citeauthoryear{{Ji}, {Oh}  \& {Masterson}}{{Ji}
  et~al.}{2019}]{ji19}
{Ji} S.,  {Oh} S.~P.,   {Masterson} P.,  2019, \mn@doi [\mnras]
  {10.1093/mnras/stz1248}, 487, 737

\bibitem[\protect\citeauthoryear{{Kim} \& {Kim}}{{Kim} \& {Kim}}{2013}]{kim}
{Kim} J.-G.,  {Kim} W.-T.,  2013, \mn@doi [\apj] {10.1088/0004-637X/779/1/48},
  779, 48

\bibitem[\protect\citeauthoryear{{Kwak} \& {Shelton}}{{Kwak} \&
  {Shelton}}{2010}]{kwak10}
{Kwak} K.,  {Shelton} R.~L.,  2010, \mn@doi [\apj]
  {10.1088/0004-637X/719/1/523}, 719, 523

\bibitem[\protect\citeauthoryear{{Mandelker}, {Nagai}, {Aung}, {Dekel},
  {Birnboim}  \& {van den Bosch}}{{Mandelker} et~al.}{2020}]{mandelker20}
{Mandelker} N.,  {Nagai} D.,  {Aung} H.,  {Dekel} A.,  {Birnboim} Y.,   {van
  den Bosch} F.~C.,  2020, \mn@doi [\mnras] {10.1093/mnras/staa812}, 494, 2641

\bibitem[\protect\citeauthoryear{{McCourt}, {Oh}, {O'Leary}  \&
  {Madigan}}{{McCourt} et~al.}{2018}]{mccourt18}
{McCourt} M.,  {Oh} S.~P.,  {O'Leary} R.,   {Madigan} A.-M.,  2018, \mn@doi
  [\mnras] {10.1093/mnras/stx2687}, \href
  {https://ui.adsabs.harvard.edu/abs/2018MNRAS.473.5407M} {473, 5407}

\bibitem[\protect\citeauthoryear{{McKee} \& {Cowie}}{{McKee} \&
  {Cowie}}{1977}]{mckee77}
{McKee} C.~F.,  {Cowie} L.~L.,  1977, \mn@doi [\apj] {10.1086/155350}, \href
  {https://ui.adsabs.harvard.edu/abs/1977ApJ...215..213M} {215, 213}

\bibitem[\protect\citeauthoryear{{Scannapieco} \& {Br{\"u}ggen}}{{Scannapieco}
  \& {Br{\"u}ggen}}{2015}]{scannapieco15}
{Scannapieco} E.,  {Br{\"u}ggen} M.,  2015, \mn@doi [\apj]
  {10.1088/0004-637X/805/2/158}, 805, 158

\bibitem[\protect\citeauthoryear{{Schneider} \& {Robertson}}{{Schneider} \&
  {Robertson}}{2017}]{schneider17}
{Schneider} E.~E.,  {Robertson} B.~E.,  2017, \mn@doi [\apj]
  {10.3847/1538-4357/834/2/144}, \href
  {https://ui.adsabs.harvard.edu/abs/2017ApJ...834..144S} {834, 144}

\bibitem[\protect\citeauthoryear{{Slavin}, {Shull}  \& {Begelman}}{{Slavin}
  et~al.}{1993}]{slavin93}
{Slavin} J.~D.,  {Shull} J.~M.,   {Begelman} M.~C.,  1993, \mn@doi [\apj]
  {10.1086/172494}, 407, 83

\bibitem[\protect\citeauthoryear{{Spitzer}}{{Spitzer}}{1962}]{spitzer}
{Spitzer} L.,  1962, Physics of Fully Ionized Gases.
Wiley-Interscience

\bibitem[\protect\citeauthoryear{{Stone}, {Tomida}, {White}  \&
  {Felker}}{{Stone} et~al.}{2020}]{Stone2020}
{Stone} J.~M.,  {Tomida} K.,  {White} C.~J.,   {Felker} K.~G.,  2020, \mn@doi
  [\apjs] {10.3847/1538-4365/ab929b}, 249, 4

\bibitem[\protect\citeauthoryear{{Tan}, {Oh}  \& {Gronke}}{{Tan}
  et~al.}{2021}]{tan21}
{Tan} B.,  {Oh} S.~P.,   {Gronke} M.,  2021, \mn@doi [\mnras]
  {10.1093/mnras/stab053}, \href
  {https://ui.adsabs.harvard.edu/abs/2021MNRAS.502.3179T} {502, 3179}

\bibitem[\protect\citeauthoryear{{Turk}, {Smith}, {Oishi}, {Skory}, {Skillman},
  {Abel}  \& {Norman}}{{Turk} et~al.}{2011}]{yt}
{Turk} M.~J.,  {Smith} B.~D.,  {Oishi} J.~S.,  {Skory} S.,  {Skillman} S.~W.,
  {Abel} T.,   {Norman} M.~L.,  2011, \mn@doi [\apjs]
  {10.1088/0067-0049/192/1/9}, 192, 9

\bibitem[\protect\citeauthoryear{{Wakker}, {Savage}, {Fox}, {Benjamin}  \&
  {Shapiro}}{{Wakker} et~al.}{2012}]{wakker12}
{Wakker} B.~P.,  {Savage} B.~D.,  {Fox} A.~J.,  {Benjamin} R.~A.,   {Shapiro}
  P.~R.,  2012, \mn@doi [\apj] {10.1088/0004-637X/749/2/157}, \href
  {https://ui.adsabs.harvard.edu/abs/2012ApJ...749..157W} {749, 157}

\bibitem[\protect\citeauthoryear{{Wiersma}, {Schaye}  \& {Smith}}{{Wiersma}
  et~al.}{2009}]{wiersma09}
{Wiersma} R. P.~C.,  {Schaye} J.,   {Smith} B.~D.,  2009, \mn@doi [\mnras]
  {10.1111/j.1365-2966.2008.14191.x}, \href
  {https://ui.adsabs.harvard.edu/abs/2009MNRAS.393...99W} {393, 99}

\makeatother
\end{thebibliography}

\bsp
\label{lastpage}
\end{document}